\documentclass[a4paper]{article}

\usepackage{INTERSPEECH2019}

\usepackage{graphicx}
\usepackage{amssymb,amsmath,bm}
\usepackage{textcomp}

\newcommand{\Table}[1]{Table \ref{tb:#1}} 
\newcommand{\Eq}[1]{Eq. (\ref{eq:#1})} 
\newcommand{\Sec}[1]{{\bf Section \ref{sec:#1}}} 
\newcommand{\Figure}[3]{\vspace{-0mm} \includegraphics[width=#1,clip]{#2.eps} \vspace{-0mm} \caption{#3} \vspace{-0mm} \label{fig:#2}}

\renewcommand{\Vec}[1]{\textrm{\boldmath $#1$}} 

\newcommand{\pt}[1]{\left(#1\right)} 
\newcommand{\x}{ \Vec{x} } 
\newcommand{\y}{ \Vec{y} } 
\newcommand{\haty}{ \Vec{\hat y} } 
  
\newcommand{\drawfig}[4]{ 
  \begin{figure}[#1]
  \begin{center}
  \Figure{#2}{#3}{#4} 
  \end{center} 
  \end{figure}
}
\graphicspath{{./figure/}}

\ninept

\title{V2S attack: building DNN-based voice conversion\\from automatic speaker verification}

\makeatletter
\def\name#1{\gdef\@name{#1\\}}
\makeatother
\name{{\em Taiki Nakamura$^{1}$, Yuki Saito$^2$, Shinnosuke Takamichi$^2$, Yusuke Ijima$^3$, and Hiroshi Saruwatari$^{2}$}}

\address{$^1$ Faculty of Engineering, The University of Tokyo, Japan. \\
$^2$ Graduate School of Information Science and Technology, The University of Tokyo, Japan. \\
$^3$ Nippon Telegraph and Telephone Corporation, Japan.
}
\email{
supikiti22@gmail.com,
\{yuuki\_saito,shinnosuke\_takamichi,hiroshi\_saruwatari\}@ipc.i.u-tokyo.ac.jp,
ijima.yusuke@lab.ntt.co.jp
}

\begin{document}
\setlength\textfloatsep{6pt}

\maketitle

\begin{abstract}
This paper presents a new voice impersonation attack using voice conversion (VC). Enrolling personal voices for automatic speaker verification (ASV) offers natural and flexible biometric authentication systems. Basically, the ASV systems do not include the users' voice data. However, if the ASV system is unexpectedly exposed and hacked by a malicious attacker, there is a risk that the attacker will use VC techniques to reproduce the enrolled user's voices. We name this the ``verification-to-synthesis (V2S) attack'' and propose VC training with the ASV and pre-trained automatic speech recognition (ASR) models and without the targeted speaker's voice data. The VC model reproduces the targeted speaker's individuality by deceiving the ASV model and restores phonetic property of an input voice by matching phonetic posteriorgrams predicted by the ASR model. The experimental evaluation compares converted voices between the proposed method that does not use the targeted speaker's voice data and the standard VC that uses the data. The experimental results demonstrate that the proposed method performs comparably to the existing VC methods that trained using a very small amount of parallel voice data.
\end{abstract}

\section{Introduction} \label{sec:intro}
Automatic speaker verification (ASV), which offers natural and flexible biometric authentication systems, has been actively studied in recent decades~\cite{najim11ivector,variani14d-vector}. Because the ASV systems identify the speaker of the input voice without using other biometrics, they are preferred for use in keyword spotting~\cite{prabhavalkar15gaincontrolkeyword} and voice search implemented in smartphones. Among the ASV systems, text-independent ASV has the potential for highly portable speaker verification. 

With deployments of ASV systems, we need to discuss the possibility of \textit{voice impersonation attack} via the ASV systems. Specifically, if a malicious attacker exposes and hacks the ASV models, voices of the enrolled speakers risk being reproduced by the attacker. Voice conversion (VC)~\cite{stylianou88,toda07_MLVC,saito18advss}, which converts voices into the targeted speaker's ones, is a possible way for this type of attack. We call this attack \textit{verification-to-synthesis (V2S) attack} that builds VC models from the pre-trained ASV model. Since ASV systems basically do not include voice data of the targeted speaker, we cannot perform the standard VC training using the targeted speaker's voice. However, since the ASV model learns the speaker's individuality, deceiving the model has some possibility of reproducing the targeted speaker's individuality by VC.  

This paper proposes a V2S attack using a VC model trained with a ASV model. In this paper, we use a ``white-boxed'' ASV model, which means the attacker knows a deep neural network (DNN) architecture and the targeted speaker's label. Since the ASV model does not use phonetic property of the input voice, training the VC model using only the ASV model will lose phonetic property of the converted voice. Therefore, we further use the automatic speech recognition (ASR) model prepared by the attacker for restoring the phonetic property. The VC model is trained by not only deceiving the ASV model but also matching the output of the ASR model (i.e., phonetic posteriorgrams~\cite{sun16}) predicted from the input and converted voices. In the experimental evaluation, we evaluate the performance (i.e., naturalness and speaker individuality of the converted voice) of the proposed V2S attack with the existing VC methods (\Sec{vc}) because their performances are the upper limit of the proposed method. The experimental results demonstrate that the proposed method performs comparably to the existing VC methods trained using a very small amount of parallel voices.

This paper is organized as follows. \Sec{vc} briefly reviews conventional
VC methods that require the targeted speaker's voice data. \Sec{v2s}
introduces the V2S attack that constructs the VC model with the ASV model and without the targeted speaker's voice. \Sec{exp} presents the experimental
evaluations. \Sec{concl} concludes this paper with a summary.

    \drawfig{t}{0.85\linewidth}{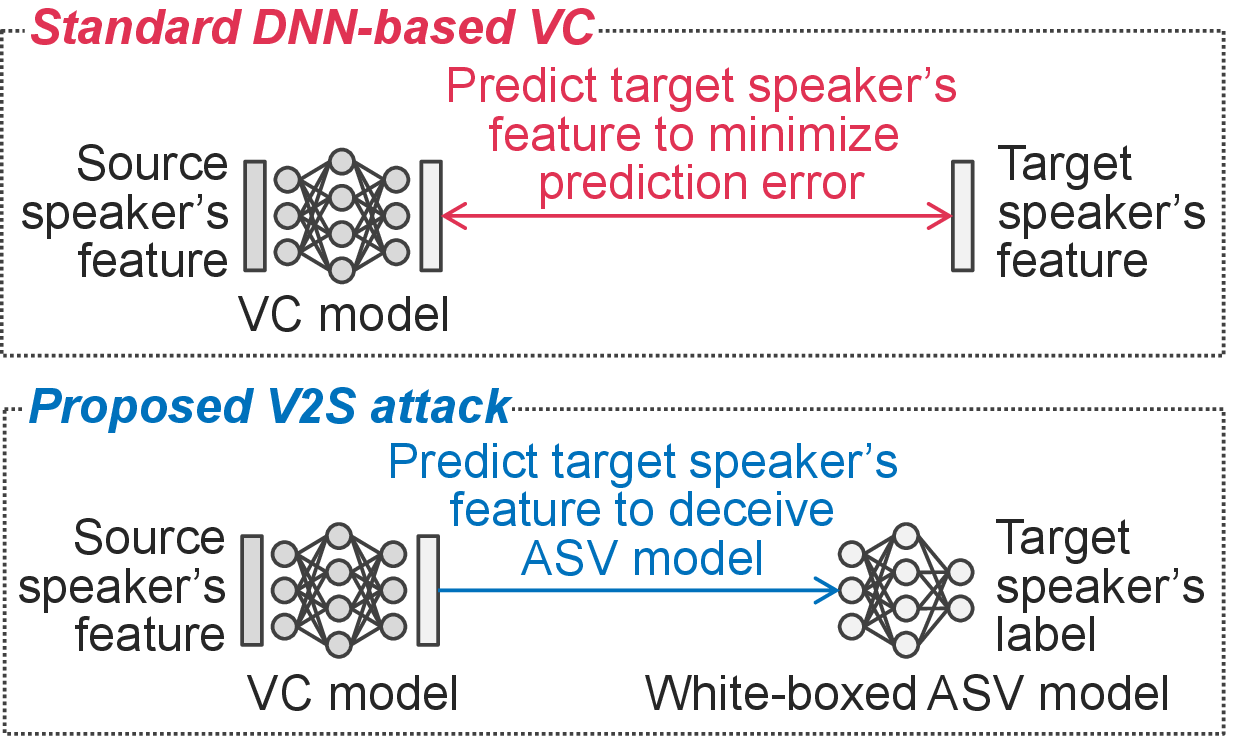}
    {Comparison of standard DNN-based voice conversion (VC) described in \Sec{vc} and proposed verification-to-synthesis (V2S) attack described in \Sec{v2s}. ASV indicates automatic speaker verification.}

\section{Building voice conversion using targeted speaker's voice} \label{sec:vc}
This section describes standard VC techniques using a targeted speaker's voice: parallel and non-parallel VC (shown in the upper half of Fig.~\ref{fig:fig/overview}). They are references to evaluate the performances of the proposed V2S attack. 

    \subsection{One-to-one parallel VC~\cite{desai09nnvc}} \label{sec:para-vc}
    Let $\Vec{G}(\cdot)$ be a VC model (a.k.a., an acoustic model), and let $\x = [\x_1^\top, \cdots, \x_t^\top, \cdots, \x_T^\top]^\top$ and $\y = [\y_1^\top, \cdots, \y_t^\top, \cdots, \y_T^\top]^\top$ be the source and targeted speakers' acoustic feature sequences extracted from a parallel speech corpus, respectively. $t$ and $T$ denote the frame index and total frame length, respectively. $\x_t$ and $\y_t$ are the source and targeted speakers' acoustic feature vectors at frame $t$, respectively. A converted acoustic feature sequence $\haty = [\haty_1^\top, \cdots, \haty_t^\top, \cdots, \haty_T^\top]^\top$ is given as $\haty = \Vec{G}(\x)$. $\Vec{G}(\cdot)$ is trained to minimize a prediction error: e.g., mean squared error (MSE) between $\y$ and $\haty$ defined as
        \begin{align}
            L_{\mathrm{MSE}}\pt{\y, \haty} = \frac{1}{T} \pt{\haty - \y}^\top\pt{\haty - \y}.
        \end{align}
    
    \subsection{One-to-many non-parallel VC} \label{sec:non-para-vc}
    In one-to-many VC, i.e., VC from a source speaker to any arbitrary speakers, $\Vec{G}(\cdot)$ is often trained using multi-speaker corpora in advance, and is adapted using the specific targeted speaker's voice data. This paper utilizes the $d$-vector-based model adaptation method~\cite{saito2018non}. The method employs another DNN that estimates a $d$-vector (i.e., DNN-based continuous speaker representation~\cite{variani14d-vector}) of the targeted speaker, and feeds it to the VC model to convert the source speaker's features into the targeted speaker's ones.

\section{V2S attack: building voice conversion without using targeted speaker's voice} \label{sec:v2s}
This section proposes a novel voice impersonation attack named a V2S attack (shown in the lower half of Fig.~\ref{fig:fig/overview}). Unlike methods described in \Sec{vc}, the VC model is trained from a white-boxed ASV model but without the targeted speaker's voice data. Besides the ASV model, we use an ASR model for the VC model training. Deceiving the ASV model helps to reproduce the targeted speaker's individuality and using the ASR model helps to restore the phonetic property of the input voice.

    \subsection{ASV model to be attacked}
    We assume that the attacked ASV system has an ASV model $\Vec{V}\pt{\cdot}$ that extracts a latent variable of the speaker identity from input speech. In this paper, we construct a $d$-vector-based ASV model~\cite{variani14d-vector} and train it to recognize one of the enrolled $S$ speakers. The $s_y$th speaker is identified by the one-hot speaker code $\Vec{l}_{y} = [l_y(1), \cdots, l_y(s), \cdots, l_y(S)]^{\top}$ whose element is defined as
    \begin{align}
        l_y(s) &=
        \begin{cases}
            1 \;\; {\rm if} \;\; s = s_y \\
            0  \; \; {\rm otherwise} 
        \end{cases} \; (1 \le s \le S),
    \end{align}
    where $s$ is the speaker index. At run time, $\Vec{V}\pt{\cdot}$ outputs a frame-level posterior probability of the specific targeted speaker. Let $\Vec{V}(\y) = [\Vec{v}_{1}^{\top}, \cdots, \Vec{v}_{t}^{\top}, \cdots, \Vec{v}_{T}^{\top}]^{\top}$ be the probability sequence. $\Vec{v}_{t} = [v_{t}(1), \cdots, v_{t}(s), \cdots, v_{t}(S)]^{\top}$ is the probability vector at frame $t$. $v_{t}(s)$ is the probability that $\y_t$ is uttered by the $s$th speaker. $\Vec{V}\pt{\cdot}$ is trained to minimize the softmax cross-entropy (SCE) loss defined as
    \begin{align}
        L_{\mathrm{SCE}}\pt{\Vec{l}_y, \Vec{V}\pt{\y} }
        & = - \frac{1}{T} \sum_{t = 1}^{T} \sum_{s = 1}^{S} l_y\pt{s} \log v_{t}\pt{s}  \label{eq:L_SCE}.
    \end{align}
    
    As described in \Sec{intro}, we set a situation in which the DNN architecture of $\Vec{V}(\cdot)$ and the targeted speaker's label
    are given. Therefore, we can estimate the difference between the targeted speaker's individuality and that of the converted acoustic features, $\haty$, as $L_{\mathrm{SCE}}\pt{\Vec{l}_y, \Vec{V}\pt{\haty}}$ that can be backpropagated to other DNNs for their training.
    
    \subsection{ASR model to restore phonetic property}
    Training the VC model to deceive the ASV model, i.e., to minimize $L_{\mathrm{SCE}}\pt{\Vec{l}_y, \Vec{V}\pt{\haty}}$, can be expected to reproduce the targeted speaker's individuality. However, it does not guarantee that the phonetic property of the input voice will be restored during the VC process. Therefore, we use the ASR model $\Vec{R}(\Vec{\cdot})$ to compute the discrepancy between the phonetic properties of the input and converted voices. As $\Vec{R}(\Vec{\cdot})$ predicts frame-level phonetic posteriorgrams of the input voice, we can estimate the discrepancy as the MSE between $\Vec{R}(\x)$ and $\Vec{R}(\haty)$, i.e., $L_{\mathrm{MSE}}\pt{\Vec{R}\pt{\x}, \Vec{R}\pt{\haty}}$, and can use it for the proposed VC model training.
    
    \subsection{Training of VC model}
    A loss function $L(\cdot)$ for the proposed VC model training is given as 
        \begin{align}
            L\pt{\x, \haty, \Vec{l}_y} = L_{\mathrm{SCE}}\pt{\Vec{l}_y, \Vec{V}\pt{\haty}} + \omega L_{\mathrm{MSE}}\pt{\Vec{R}\pt{\x}, \Vec{R}\pt{\haty}}, \label{eq:l_v2s}
        \end{align}
    where $\omega$ is a hyperparameter that controls the weight of the second term. Note that Eq.~(\ref{eq:l_v2s}) does not include the targeted speaker's acoustic features $\y$ and therefore we can construct the VC model without using them. Figure \ref{fig:fig/architecture} illustrates a conceptual diagram of the proposed V2S attack.
    \drawfig{t}{0.96\linewidth}{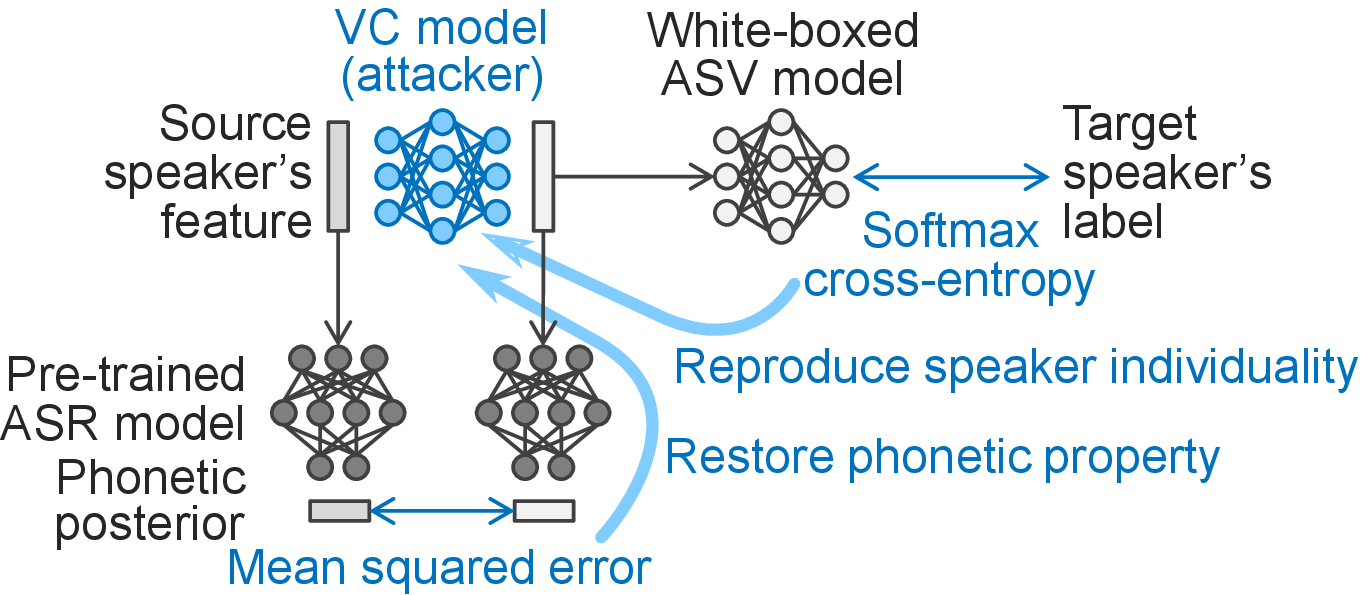}
    {Conceptual diagram of proposed V2S attack.}    

    \subsection{Discussion}
    The V2S attack is regarded as one kind of \textit{voice spoofing attack}~\cite{wu13_spoofing}. The aim is the same: masquerading oneself as another targeted speaker. So far, studies on voice spoofing attacks have focused on spoofing attacks (spoofing by synthesized voices)~\cite{wu13_spoofing} and replay attacks (spoofing by replayed voices)~\cite{lindberg99replayattack}. The V2S attack follows the former and tries to reproduce the targeted speaker's voice from the ASV model. A similar idea was used in Saito et al.'s work~\cite{saito18advss} that incorporated a voice anti-spoofing (i.e., a discriminative model to detect spoofing attacks) into training of a VC model for reproducing fine structures of the synthesized voice.
    
    From the above perspectives, our approach is close to that of Kinnunen et al.~\cite{kinnunen19speakerattack}. They proposed algorithms to select an utterance from public-domain corpora to deceive the ASV model. On the other hand, we aim to build the VC model for synthesizing every utterance of the targeted speaker.
    
    An adversarial attack (e.g., for automatic speaker recognition~\cite{yukara18audioadversarial} and image classification~\cite{goodfellow14adversarialexample}) is the common attack for pre-trained recognition models. The purpose of the V2S attack is completely different from that of the adversarial attack: the adversarial attack aims to let the discriminator misclassify the input sample, but the V2S attack aims to reproduce attributes of the targeted sample (e.g., speaker individuality) from the discriminator.
    
    From the above perspectives, a classifier-to-generator attack~\cite{kusano18classifiertogenerator} and a membership inference attack~\cite{hisamoto19membershipinference} are approaches related to our V2S attack. They attack a pre-trained model to estimate the training data distribution or the data themselves from the model. Using these attacks is expected to improve naturalness in our converted voice, e.g., estimating utterances close to the training data of the ASV model will help deception by the VC model.
    
    The proposed training algorithm does not use pre-stored speakers' voice data, unlike one-to-many non-parallel VC~\cite{saito2018non} and Kinnunen et al.'s work~\cite{kinnunen19speakerattack}. One of our future directions is to use the pre-stored speakers' voice data to improve speaker individuality of the converted voice (e.g., using transfer learning~\cite{jia18}). Also, integrating high-fidelity waveform generation~\cite{oord16wavenet} and using end-to-end ASV and ASR models~\cite{zhang17end2endasv,zhang17end2endasr} is remaining challenges for the V2S attack in more realistic situations.
    
    In this paper, the VC model is trained by attacking an white-boxed ASV model. A more realistic situation is an attack against the black-boxed ASV model, i.e., features, DNN architectures, and targeted speaker's label are unknown for the attacker. For instance, black-box optimization based on generative adversarial networks~\cite{goodfellow14gan,fu19} or reinforcement learning~\cite{koizumi17} can be introduced to the proposed V2S attack. Performances of the white-boxed case described in this paper will be the reference for the black-boxed case. 

\section{Experimental evaluation}\label{sec:exp}
    \subsection{Experimental setup}
    Acoustic features used in VC, ASV, and ASR were 39-dimensional (1st-through-39th) mel-cepstral coefficients and their delta features. The STRAIGHT~\cite{kawahara99} vocoder systems were used for the feature extraction. The frame shift was 5~ms. We set two types of conversion: male-to-male and male-to-female. One male was the source speaker, and two males and two females were the targeted speakers. We conducted evaluation for each pair of source and targeted speakers using their 25 parallel voices as the evaluation data.
    
    The VC model for the V2S attack was a Feed-Forward neural network with $78$ input units, $\left\{256, 128\right\}$ ReLU~\cite{NairH10relu} hidden units, and $78$ output units. The ASV model was a Feed-Forward neural network with $78$ input units, $\left\{1024\times3, 8\times1\right\}$ sigmoid hidden units, and $260$ output units. The ASR model was a Feed-Forward neural network with $78$ input units, $1024\times4$ sigmoid hidden units, and $56$ output units ($56$ is the number of phonemes). Utterances by $260$ Japanese speakers ($130$ males and $130$ females) were used for training the ASV and ASR models. The four targeted speakers (two males and two females) were in the 260 speakers since we performed the proposed V2S attack towards the white-boxed ASV model. AdaGrad~\cite{duchi11adagrad} with the learning rate setting to $0.1$ was used for training the VC model. The VC model training was performed with 25 epochs using 200 utterances of the source speaker. The weight $\omega$ in \Eq{l_v2s} was set to 0.01. In the V2S attack, the spectral parameters were converted by the VC model. The 0th mel-cepstral coefficients and band-aperiodicity were not converted, i.e., the original speaker's features were directly used for waveform generation. For $F_0$ conversion, the targeted speaker's $F_0$ is not observed in the V2S attack. This problem is not solved in this paper. Therefore, we calculated $F_0$ statistics (mean and variance) from the targeted speaker's voice data, and performed linear conversion of $F_0$~\cite{toda07_MLVC}.

    In one-to-one parallel VC, we trained a DNN for spectral conversion. The DNN architecture was the same as that for the VC model of the V2S attack. The number of training data is discussed in the following section. The DNN was trained with 25 epochs. The $F_0$ and band-aperiodicity conversions are the same as those in the V2S attack. In one-to-many non-parallel VC, the DNN architectures, training data, and other hyperparameters are the same as those in Saito et al.'s work~\cite{saito2018non} that used variational autoencoders~\cite{kingma2013vae} conditioned by phonetic posteriorgrams~\cite{sun16} and $d$-vectors~\cite{variani14d-vector} as the VC model.
    
    \subsection{Experimental results}
    We compared converted voices of the proposed V2S attack with those of the existing VC methods. We conducted preference AB tests on the naturalness of the converted voice. Methods to be compared were listed as follows.
        \begin{itemize}
            \item \textbf{ParaVC}: parallel VC (\Sec{para-vc}) trained with 5, 10, or 30 utterances
            \item \textbf{NonparaVC}: Non-parallel VC (\Sec{non-para-vc})
            \item \textbf{V2S}: proposed V2S attack (\Sec{v2s})
        \end{itemize}
    We presented a pair of converted voices in random order and had listeners select the voice sample that sounded more natural. Similarly, preference XAB tests on the speaker individuality were conducted using the natural voices of the targeted speakers as reference samples ``X.'' These tests were done in our crowdsourcing evaluation system. Forty listeners participated in each evaluation, and each listener evaluated 10 samples randomly extracted from the evaluation data. The total number of listeners was 2 (AB or XAB) $\times$ 2 (male-to-male or male-to-female) $\times$ 4 (reference methods) $\times$ 40 (listeners) $=$ 640.

    \Table{subj_ab_m2m} and \Table{subj_ab_m2f} list results of the preference AB tests for male-to-male and male-to-female conversion, respectively. Similarly, \Table{subj_xab_m2m} and \Table{subj_xab_m2f} list those of the preference XAB tests for male-to-male and male-to-female conversion, respectively. From \Table{subj_ab_m2m} and \Table{subj_ab_m2f}, the proposed V2S attack has lower quality than ``NonparaVC.'' On the other hand, it has comparable or superior quality to ``ParaVC (5 utts)'' for both male-to-male and male-to-female conversion. From \Table{subj_xab_m2m} and \Table{subj_xab_m2f}, the V2S attack performed worse in terms of speaker individuality than most settings of the existing VC, but it performed comparably to ``ParaVC (5 utts)'' for male-to-male conversion. From these results, the proposed V2S attack is potentially comparable to the standard parallel VC with a very small amount of training data.
    
	  \begin{table}[t]
      \centering
      \caption{Results of preference AB tests on naturalness (male-to-male). \textbf{Bold} indicates the method preferred more with $p$-value $< 0.05$.}
      \label{tb:subj_ab_m2m}
      \begin{tabular}{|c|cc|c|}
      \hline A & Scores & $p$-value & B \\
      \hline
      \hline ParaVC ( 5~utts) & 0.388 vs. \textbf{0.612} & 1.221$\times10^{-10}$ & V2S \\
      \hline ParaVC (10~utts) & 0.475 vs. 0.525 & 0.158  		& V2S \\ 
      \hline ParaVC (30~utts) & 0.458 vs. \textbf{0.542} & 0.016 & V2S \\
      \hline NonparaVC        & \textbf{0.598} vs. 0.402 & 2.694$\times10^{-8}$ & V2S \\
      \hline
      \end{tabular}
      \end{table}

	  \begin{table}[t]
      \centering
      \caption{Results of preference AB tests on naturalness (male-to-female). \textbf{Bold} indicates the method preferred more with $p$-value $< 0.05$.}
      \label{tb:subj_ab_m2f}
      \begin{tabular}{|c|cc|c|}
      \hline A & Scores & $p$-value & B \\
      \hline
      \hline ParaVC ( 5~utts) & 0.490 vs. 0.510 & 0.572 & V2S \\
      \hline ParaVC (10~utts) & \textbf{0.593} vs. 0.407 & 1.365$\times10^{-7}$  		& V2S \\ 
      \hline ParaVC (30~utts) & \textbf{0.610} vs. 0.390 & 3.174$\times10^{-10}$ & V2S \\
      \hline NonparaVC        & \textbf{0.538} vs. 0.462 & 0.034 & V2S \\
      \hline
      \end{tabular}
      \end{table}

	  \begin{table}[t]
      \centering
      \caption{Results of preference XAB tests on speaker individuality (male-to-male). \textbf{Bold} indicates the method preferred more with $p$-value $< 0.05$.}
      \label{tb:subj_xab_m2m}
      \begin{tabular}{|c|cc|c|}
      \hline A & Scores & $p$-value & B \\
      \hline
      \hline ParaVC ( 5~utts) & 0.530 vs. 0.470 & 0.090 & V2S \\
      \hline ParaVC (10~utts) & \textbf{0.615} vs. 0.385 & $<10^{-10}$ & V2S \\ 
      \hline ParaVC (30~utts) & \textbf{0.675} vs. 0.325 & $<10^{-10}$ & V2S \\
      \hline NonparaVC        & \textbf{0.660} vs. 0.340 & $<10^{-10}$ & V2S \\
      \hline
      \end{tabular}
      \end{table}

	  \begin{table}[t]
      \centering
      \caption{Results of preference XAB tests on speaker individuality (male-to-female). \textbf{Bold} indicates the method preferred more with $p$-value $< 0.05$.}
      \label{tb:subj_xab_m2f}
      \begin{tabular}{|c|cc|c|}
      \hline A & Scores & $p$-value & B \\
      \hline
      \hline ParaVC ( 5~utts) & \textbf{0.585} vs. 0.415 & 1.324$\times10^{-6}$ & V2S \\
      \hline ParaVC (10~utts) & \textbf{0.713} vs. 0.287 & $<10^{-10}$  		& V2S \\ 
      \hline ParaVC (30~utts) & \textbf{0.705} vs. 0.295 & $<10^{-10}$ & V2S \\
      \hline NonparaVC        & \textbf{0.588} vs. 0.412 & $<10^{-10}$ & V2S \\
      \hline
      \end{tabular}
      \end{table}

\section{Conclusion}\label{sec:concl}
This paper presents a new voice impersonation attack using voice conversion (VC), named the verification-to-synthesis (V2S) attack. The VC model was trained to deceive the white-boxed automatic speaker verification (ASV) model for reproducing the targeted speaker's individuality and to restore phonetic property of the input voice by using pre-trained automatic speech recognition (ASR) model. The experimental results indicated that the proposed V2S attack can synthesize voice that has naturalness and speaker individuality comparable to an existing parallel VC with a very small amount of training data. In future work, we will evaluate the V2S attack that uses pre-stored speakers' voice data and investigate the dependency of the input speaker in our method. 

\textbf{Acknowledgements:}
Part of this work was supported by the SECOM Science and Technology Foundation.

\bibliographystyle{IEEEbib}
\bibliography{template_camera_ready}

\begin{thebibliography}{10}
\providecommand{\url}[1]{#1}
\csname url@samestyle\endcsname
\providecommand{\newblock}{\relax}
\providecommand{\bibinfo}[2]{#2}
\providecommand{\BIBentrySTDinterwordspacing}{\spaceskip=0pt\relax}
\providecommand{\BIBentryALTinterwordstretchfactor}{4}
\providecommand{\BIBentryALTinterwordspacing}{\spaceskip=\fontdimen2\font plus
\BIBentryALTinterwordstretchfactor\fontdimen3\font minus
  \fontdimen4\font\relax}
\providecommand{\BIBforeignlanguage}[2]{{%
\expandafter\ifx\csname l@#1\endcsname\relax
\typeout{** WARNING: IEEEtran.bst: No hyphenation pattern has been}%
\typeout{** loaded for the language `#1'. Using the pattern for}%
\typeout{** the default language instead.}%
\else
\language=\csname l@#1\endcsname
\fi
#2}}
\providecommand{\BIBdecl}{\relax}
\BIBdecl

\bibitem{najim11ivector}
N.~Dehak, P.~J. Kenny, R.~Dehak, P.~Dumouchel, and P.~Ouellet, ``Front-end
  factor analysis for speaker verification,'' \emph{IEEE Transactions on Audio,
  Speech, and Language Processing}, vol.~14, no.~4, pp. 788--798, 2011.

\bibitem{variani14d-vector}
E.~Variani, X.~Lei, E.~McDermott, I.~L. Moreno, and J.~Gonzalez-Dominguez,
  ``Deep neural networks for small footprint text-dependent speaker
  verification,'' in \emph{Proc. ICASSP}, Florence, Italy, May 2014, pp.
  4080--4084.

\bibitem{prabhavalkar15gaincontrolkeyword}
R.~Prabhavalkar, R.~Alvarez, C.~Parada, P.~Nakkiran, and T.~Sainath,
  ``Automatic gain control and multi-style training for robust small-footprint
  keyword spotting with deep neural networks,'' in \emph{Proc. ICASSP},
  Brisbane, Australia, Apr. 2015, pp. 4704--4708.

\bibitem{stylianou88}
Y.~Stylianou, O.~Capp\'{e}, and E.~Moulines, ``Continuous probabilistic
  transform for voice conversion,'' \emph{IEEE Transactions on Speech and Audio
  Processing}, vol.~6, no.~2, pp. 131--142, Mar. 1998.

\bibitem{toda07_MLVC}
T.~Toda, A.~W. Black, and K.~Tokuda, ``Voice conversion based on maximum
  likelihood estimation of spectral parameter trajectory,'' \emph{IEEE
  Transactions on Audio, Speech, and Language Processing}, vol.~15, no.~8, pp.
  2222--2235, 2007.

\bibitem{saito18advss}
Y.~Saito, S.~Takamichi, and H.~Saruwatari, ``Statistical parametric speech
  synthesis incorporating generative adversarial networks,'' \emph{IEEE/ACM
  Transactions on Audio, Speech, and Language Processing}, vol.~26, no.~1, pp.
  755--767, Jun. 2018.

\bibitem{sun16}
L.~Sun, K.~Li, H.~Wang, S.~Kang, and H.~Meng, ``Phonetic posteriorgrams for
  many-to-one voice conversion without parallel data training,'' in \emph{Proc.
  ICME}, Seattle, U.S.A., Jul. 2016.

\bibitem{desai09nnvc}
S.~Desai, E.~V. Raghavendra, B.~Yegnanarayana, A.~W. Black, and K.~Prahallad,
  ``Voice conversion using artificial neural networks,'' in \emph{Proc.
  ICASSP}, Taipei, Taiwan, Apr. 2009, pp. 3893--3896.

\bibitem{saito2018non}
Y.~Saito, Y.~Ijima, K.~Nishida, and S.~Takamichi, ``Non-parallel voice
  conversion using variational autoencoders conditioned by phonetic
  posteriorgrams and d-vectors,'' in \emph{Proc. ICASSP}, Calgary, Canada, Apr.
  2018, pp. 5274--5278.

\bibitem{wu13_spoofing}
Z.~Wu, X.~Xiao, E.~S. Chng, and H.~Li, ``Synthetic speech detection using
  temporal modulation feature,'' in \emph{Proc. ICASSP}, Vancouver, Canada,
  May. 2013, pp. 7234--7238.

\bibitem{lindberg99replayattack}
J.~Lindberg and M.~Blomberg, ``Vulnerability in speaker verification - a study
  of technical impostor techniques,'' in \emph{Proc. EUROSPEECH}, Budapest,
  Hungary, Mar. 1999, pp. 1211--1214.

\bibitem{kinnunen19speakerattack}
T.~Kinnunen, R.~G. Hautamäki, V.~Vestman, and M.~Sahidullah, ``Can we use
  speaker recognition technology to attack itself? enhancing mimicry attacks
  using automatic target speaker selection,'' in \emph{Proc. ICASSP}, Brighton,
  United Kingdom, May 2019.

\bibitem{yukara18audioadversarial}
H.~Yakura and J.~Sakuma, ``Robust audio adversarial example for a physical
  attack,'' \emph{arXiv:1810.11793}, 2018.

\bibitem{goodfellow14adversarialexample}
I.~J. Goodfellow, J.~Shlens, and C.~Szegedy, ``Explaining and harnessing
  adversarial examples,'' \emph{arXiv:1412.6572}, 2014.

\bibitem{kusano18classifiertogenerator}
\BIBentryALTinterwordspacing
K.~Kusano and J.~Sakuma, ``Classifier-to-generator attack: Estimation of
  training data distribution from classifier,'' 2018. [Online]. Available:
  \url{https://openreview.net/forum?id=SJOl4DlCZ}
\BIBentrySTDinterwordspacing

\bibitem{hisamoto19membershipinference}
S.~Hisamoto, M.~Post, and K.~Duh, ``Membership inference attacks on
  sequence-to-sequence models,'' \emph{arXiv:1904.05506}, 2019.

\bibitem{jia18}
Y.~Jia, Y.~Zhang, R.~J. Weiss, Q.~Wang, J.~Shen, F.~Ren, Z.~Chen, P.~Nguyen,
  R.~Pang, I.~L. Moreno, and Y.~Wu, ``Transfer learning from speaker
  verification to multispeaker text-to-speech synthesis,'' vol. abs/1806.04558,
  2018.

\bibitem{oord16wavenet}
A.~v.~d. Oord, S.~Dieleman, H.~Zen, K.~Simonyan, O.~Vinyals, A.~Graves,
  N.~Kalchbrenner, A.~W. Senior, and K.~Kavukcuoglu, ``Wave{N}et: {A}
  generative model for raw audio,'' vol. abs/1609.03499, 2016.

\bibitem{zhang17end2endasv}
C.~Zhang and K.~Koishida, ``End-to-end text-independent speaker verification
  with flexibility in utterance duration,'' in \emph{Proc. ASRU}, Okinawa,
  Japan, Dec. 2017, pp. 584--590.

\bibitem{zhang17end2endasr}
Y.~Zhang, M.~Pezeshki, P.~Brakel, S.~Zhang, C.~L.~Y. Bengio, and A.~Courville,
  ``Towards end-to-end speech recognition with deep convolutional neural
  networks,'' \emph{arXiv}, vol. abs/1701.02720, 2017.

\bibitem{goodfellow14gan}
I.~Goodfellow, J.~Pouget-Abadie, M.~Mirza, B.~Xu, D.~WardeFarley, S.~Ozair,
  A.~Courville, and Y.~Bengio, ``Generative adversarial nets,'' in \emph{Proc.
  NIPS}, Montreal, Canada, Dec. 2014, pp. 2672--2680.

\bibitem{fu19}
S.-W. Fu, C.-F. Liao, Y.~Tsao, and S.-D. Lin, ``Metric{GAN}: {G}enerative
  adversarial networks based black-box metric scores optimization for speech
  enhancement,'' vol. abs/1905.04874, 2019.

\bibitem{koizumi17}
Y.~Koizumi, K.~Niwa, Y.~Hioka, K.~Kobayashi, and Y.~Haneda, ``{DNN}-based
  source enhancement self-optimized by reinforcement learning using sound
  quality measurements,'' in \emph{Proc. ICASSP}, Orleans, U.S.A., Mar. 2017,
  pp. 81--85.

\bibitem{kawahara99}
H.~Kawahara, I.~Masuda-Katsuse, and A.~D. Cheveigne, ``Restructuring speech
  representations using a pitch-adaptive time-frequency smoothing and an
  instantaneous-frequency-based {F}0 extraction: Possible role of a repetitive
  structure in sounds,'' \emph{Speech Communication}, vol.~27, no. 3--4, pp.
  187--207, 1999.

\bibitem{NairH10relu}
V.~Nair and G.~E. Hinton, ``Rectified linear units improve restricted boltzmann
  machines,'' in \emph{Proc. ICML}, Haifa, Israel, June 2010, pp. 807--814.

\bibitem{duchi11adagrad}
J.~Duchi, E.~Hazan, and Y.~Singer, ``Adaptive subgradient methods for online
  learning and stochastic optimization,'' \emph{Journal of Machine Learning
  Research}, vol.~12, pp. 2121--2159, Jul. 2011.

\bibitem{kingma2013vae}
D.~P. Kingma and M.~Welling, ``Auto-encoding variational bayes,'' \emph{arXiv},
  vol. abs/1312.6114, 2013.

\end{thebibliography}

\end{document}